\documentclass[final,authoryear,5p,times,twocolumn]{elsarticle}

\usepackage{graphicx}

\usepackage{amssymb,amsmath}

\journal{Planetary and Space Science}

\begin{document}

\begin{frontmatter}



\title{PROPERTIES OF THE LUNAR EXOSPHERE DURING THE PERSEID 2009 METEOR SHOWER}


\author[a1]{Berezhnoy A.A.}
\address[a1]{Sternberg Astronomical Institute, Moscow State University,
Universitetskij pr., 13, Moscow, 119991 Russia}

\author[a2]{Churyumov K.I.}
\address[a2]{Astronomical Observatory, Kyiv Taras Shevchenko National
University, Observatorna Street, 3, Kyiv, 04053 Ukraine}

\author[a2]{Kleshchenok V.V.}

\author[a1]{Kozlova E.A.}

\author[a5]{Mangano V.}
\address[a5]{INAF-IAPS (Institute of Astrophysics and Planetology from Space),
via del Fosso del Cavaliere 100, 00133 Rome, Italy}

\author[a6]{Pakhomov Yu.V.}
\address[a6]{Institute of Astronomy, Russian Academy of Sciences, Pyatnitskaya
Street 48, Moscow, 119017 Russia}

\author[a2]{Ponomarenko V.O.}

\author[a1]{Shevchenko V.V}

\author[a9]{Velikodsky Yu.I.}
\address[a9]{Institute of Astronomy, Kharkiv National University, Sumska Street
35, Kharkiv, 61022 Ukraine}

\begin{abstract}
Influence of the meteoroid bombardment on properties of the lunar exosphere has
been confirmed.
Quick increase in the zenith column density of Na atoms above the lunar north
pole on August 13, 2009
at 0--1 UT is detected and explained by numerous collisions of relatively small
Perseid meteoroids   
($<$1~kg) with the Moon during maximum of the Perseid meteor shower. New
stringent upper limits of   
the column densities for Ca, Ba, and Ti atoms in the lunar exosphere are
obtained as
$5{\times}10^7$, $2.2{\times}10^6$, and $6.9{\times}10^8$~cm$^{-2}$,
respectively. 
It is found that the content of impact-produced Ca and Al atoms in the
lunar exosphere is depleted as compared to that of Na atoms.

\end{abstract}


\end{frontmatter}


\section{Introduction}
\label{Introduction}

Spectral lines of atoms of sodium and potassium were discovered in the lunar
exosphere by Potter and Morgan (1988).
The emission lines of neutral sodium were detected at distances of
about 5 lunar radii from the center of the Moon on the sunward
side and much fainter emission was detected at distances up to
about 15--20 lunar radii on the antisunward side. The typical
velocity of Na atoms in the extended lunar coma is about 2 km/s
suggesting a high-energy source, such as impact vaporization of
the regolith (Mendillo et al., 1991). The distribution of sodium in
the lunar exosphere depends on solar zenith angle, suggesting that
most sodium atoms are liberated from the lunar surface by solar
photons or by solar wind impact, in contrast to a source driven by
uniform micrometeoroid bombardment (Flynn and Mendillo,
1993; Mendillo et al., 1993).

Properties of the extended lunar sodium exosphere have been
explained by a 15\% contribution of sporadic micrometeoroid
impact vaporization occurring uniformly over the lunar surface
and an 85\% contribution of photon-induced desorption dependent
on solar zenith angle over the sunlit hemisphere (Mendillo et al.,
1999). In particular, impact vaporization caused by collisions of
sporadic meteoroids with the Moon may account for up to 50\% of
exospheric Na atoms over the terminator and poles (Sarantos et al.,
2010). Increase in the content of sodium atoms in the lunar
exosphere is expected during activity of main meteor showers
due to increasing intensity of meteoroid bombardment. For
example, a small increase in temperature and column density of
Na atoms in the lunar exosphere was detected during Leonid 1995
and 1997 showers (Hunten et al., 1998; Verani et al., 1998), but no
similar effects were detected during Geminid 1999 and Quadrantid
1999 meteor showers (Barbieri et al., 2001; Verani et al., 2001).
A bright Na spot in the lunar orbit was detected after maximum
of the Leonid 1998 meteor shower (Smith et al., 1999). Thus,
meteoroid impacts may lead to the production of Na atoms which
are able to escape the lunar exosphere.

\begin{table*}
\caption{Parameters of spectral observations of Na atoms in the lunar exosphere
on August 12--14, 2009. The values of $N_{zen0}$ are calculated at the assumed
temperature of 3000~K and averaged from analysis of both Na (5890 and 5896~\AA)
lines. The accuracies of the values of observed intensities $I_{obs}$ and the
surface zenith column densities $N_{zen0}$ are given at the 3 sigma level.}
\small
\begin{tabular}{lllllll}
\hline
Time of observations
&
Illuminated
&
Distance from the
&
Position angle
&
Intensity of
&
Intensity of
&
$N_{zen0}$(Na)  \\

UT
&
fraction, (\%)
&
surface, (km)
&
(deg)
&
Na D2 line, (R)
&
Na D1 line, (R)
&
(cm$^2$)\\

\hline

Aug.12, 23:13--23:43      &58.8&90
&18.9&146$\pm$6&68$\pm$6&(8.2$\pm$0.5)${\times}10^8$\\
Aug.12, 23:54--Aug.13,
0:24&58.5&270&18.8&159$\pm$3&77$\pm$3&(1.23$\pm$0.04)${\times}10^9$\\
Aug.13, 0:43--1:13       
&58.2&455&18.7&137$\pm$3&64$\pm$3&(1.33$\pm$0.04)${\times}10^9$\\
Aug.13, 23:22--23:52      &48  &90
&15.2&152$\pm$3&74$\pm$3&(8.5$\pm$0.3)${\times}10^8$\\
Aug.13, 23:53--Aug.14,
0:23&47.7&270&15.1&136$\pm$3&66$\pm$3&(1.00$\pm$0.04)${\times}10^9$\\
Aug.14, 0:26--0:56       &47.4&455&15.0&
89$\pm$3&45$\pm$3&(8.7$\pm$0.5)${\times}10^8$\\
\hline
\end{tabular}
\end{table*}

Several other authors reported sudden changes in the properties of the sodium
lunar exosphere which may be associated with
impacts of meteoroids. Hunten et al. (1991) detected an increase in
the column density of Na atoms at 801 south latitude in the lunar
exosphere of about 60\% on October 14, 1990 as compared to the
observations of October 12 and 13, 1990 while measurements at
the equator showed no substantial change. These results were
explained by the action of an unknown low-speed meteor shower.
Similar quick changes in the column density of Na atoms above the
north pole of the Moon on September 18--19, 1995, were reported
and explained by impacts of numerous small low-speed meteoroids by Sprague et
al. (1998). Sudden significant changes in Na
temperature were detected during observations of the lunar poles
on April 19 and May 10, 1998, which were interpreted as possible
impacts of meteoroids (Sprague et al., 2012). These possible
impacts may be associated with Lyrid (maximum on April 22)
and $\eta$-Aquarid (maximum on May 6) meteor showers. Thus, the
column density of Na atoms in the lunar exosphere at the poles
varies significantly, but the nature of this variability remains
poorly understood.

The atoms of refractory elements, such as Ca, Mg, Al, and Fe,
have not been detected in the lunar exosphere yet, while Mg and
Ca atoms in the Hermean exosphere have, and their presence is
explained by the action of high-energy mechanisms such as solar
wind sputtering (Sarantos et al., 2011; Burger et al., 2012). Several
attempts to detect atoms of refractory elements in the lunar
exosphere have been performed (Flynn and Stern, 1996; Stern
et al., 1997; Halekas et al., 2013; Cook et al., 2013), but all these
observations were performed in the absence of main meteor
showers. For this reason, meteoroid impacts were not considered
in these papers as a source of the lunar exosphere. The search for
lines of refractory elements above the north pole of the Moon during
the Perseid 2009 meteor shower was performed by Churyumov et al.
(2012), and preliminary results of these observations were reported.
The mass of impacted Perseid meteoroids was estimated as 15 kg.
The upper limits of Ca, Ba, and Ti atoms in the lunar exosphere were
estimated using a simple barometric model as $1.6{\times}10^7$,
$7.4{\times}10^5$,
and $1.2{\times}10^7$cm$^{-2}$, respectively. However, in this paper we use the
more appropriate model of Chamberlain (1963) for the study of
exospheric atoms. In addition, we also include a geometric factor in
the data analysis, and re-analyze original observational data against
the early paper of Churyumov et al. (2012).

\section{Observations of the lunar exosphere and data reduction}

Spectroscopic observations of NaI D1 (5895.9 \AA) and D2
(5890.0 \AA) resonance lines in the lunar exosphere were performed
on August 12/13 and 13/14, 2009, during maximum of the Perseid
meteor shower using the echelle spectrograph MMCS (Multi Mode
Cassegrain Spectrometer) of the 2-m Zeiss telescope (Terskol
branch of Institute of Astronomy of Russian Academy of Sciences,
Kabardino-Balkaria, Russia). The slit of the spectrograph has a
height of 1000 and a width of 200 . Using a 1245$\times$1152 pixel CCD, 31
spectral orders in the range from 3720 to 7526 \AA\ were recorded.
The spectrograph resolution was R=13\,500; the signal-to-noise
ratio of the obtained spectra was about 50 at the position of NaI D2
line. Six echelle spectra were recorded at the distances of 50$''$,
150$''$, and 250$''$ (90, 270, and 455 km, respectively) from the lunar
limb above the north pole which was bombarded by Perseid's
meteoroids (see Table 1). The exposure time of each spectrum $\tau_{obs}$
was 1800 s.

The echelle package of the MIDAS software system was used to
process the spectroscopic data reduction: remove the cosmic rays,
detect and extract the echelle orders, wavelength calibrate using
the spectrum of a standard Fe--Ar lamp, and flux calibrate using
the standard star HD~214923. At the end of data reduction processing we obtain
the spectra in absolute fluxes. At this stage, the data
contained both the spectra of the lunar exosphere and the solar
spectra reflected from the lunar surface and scattered in the
Earth's atmosphere. To finally extract the contribution of the lunar
exosphere we use the solar spectrum taken as the spectrum of
daytime scattered light (see Fig. 1).

The spectral transparency of Earth's atmosphere at 600~nm
was taken as 88\% at 451 in accordance with Tug (1977). In Table 1,
parameters of the performed observations are listed: the time of
observations, the altitude of the slit above the lunar surface, the
position angle of the observed point with respect to the direction
of the north pole of the Moon, and the intensity of sodium
resonance lines. We notice that: (1) the brightness of Na lines at
270 and 455 km from the limb is 109\% and 93\% versus that at
90 km on August 12/13, 2009 (see Fig. 1); (2) the brightness of Na
lines at 270 and 455 km from the limb is 89\% and 58\% versus that
at 90 km on August 13/14, 2009; (3) the brightness of the reflected
solar spectrum at 270 and 455 km versus that at 90 km is 78\% and
56\% on August 12/13 and 73\% and 52\% on August 13/14, 2009,
respectively. Similar behavior of the intensity of the scattered light
in the vicinities of the Moon means that transparency of Earth's
atmosphere was almost the same during both observational
nights. More detailed information about properties of Na exosphere can be
obtained from estimates of temperature and column
density of Na atoms.

\section{The properties of the lunar exosphere}

Let us define the observed point as the nearest point at the line of
sight to the lunar surface. According to Eq. (6.2.3) of Chamberlain
and Hunten (1987) in approximation of optically thin atmosphere
the line-of-sight column abundances $N_{LOS}$ at the given altitude $h$ of
the observed point are calculated as

\begin{equation}
N_{LOS}(h) = 4{\times}10^6 \frac{\pi I}{g},
\end{equation}

where $4\pi I$ is the emission of atoms for the studied elements (in
Rayleighs) measured at the altitude $h$ and $g$ is the emission rate factor
in photons $atom^{-1}s^{-1}$ (see Table 2).

\begin{figure*}
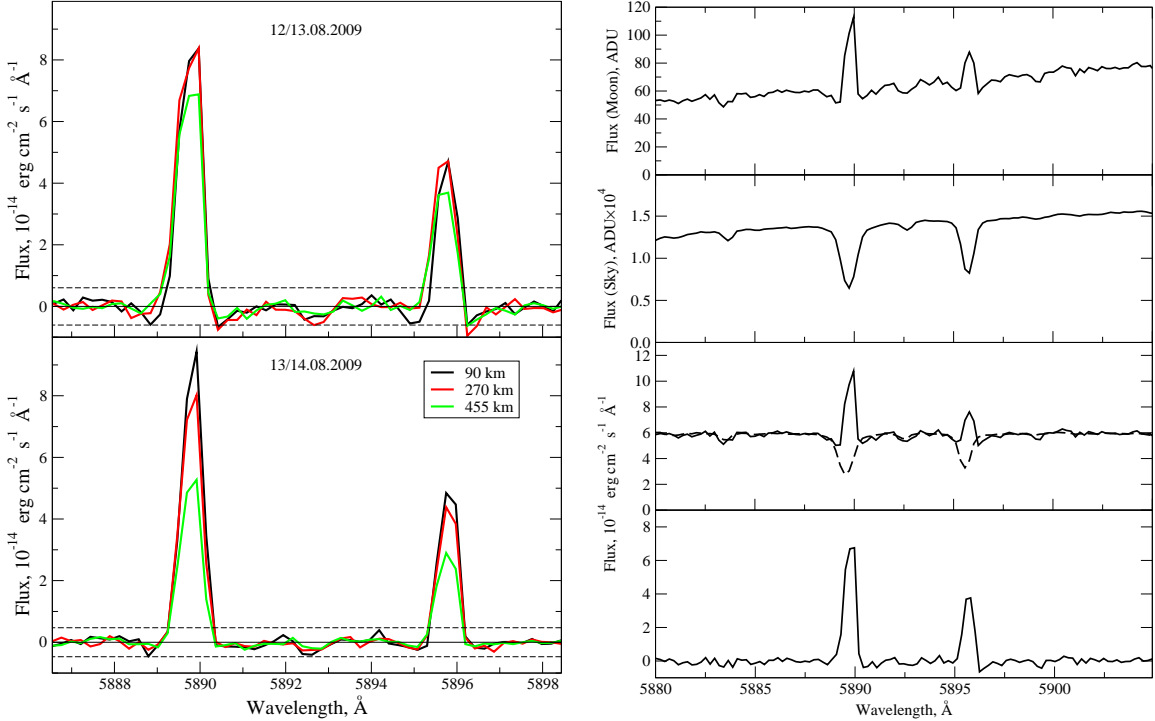

\begin{minipage}{\textwidth}
\centering
\begin{tabular}{cc}
\includegraphics[width=0.4\textwidth]{fig1a.eps} &
\includegraphics[width=0.4\textwidth]{fig1b.eps} 
\end{tabular}
\end{minipage}
\caption{Spectra of Na lines in the lunar exosphere recorded on August 12/13 and
August 13/14, 2009 (left). On the right: the sequence of data processing for the
spectrum, recorded on August 13, 0:43--1:13 UT, is presented. From top to
bottom: raw spectrum of the vicinities of the Moon in units of ADU (analog to
digital unit), raw daytime spectrum of the sky (equivalent of the solar
spectrum), calibrated spectra of the vicinities of the Moon and daytime sky
spectrum (scaled to the first), and extracted spectrum of the lunar exosphere.}
\end{figure*}

To theoretically model the properties of non-stationary exospheres, advanced
Monte Carlo technique can be applied (see, for
example, Goldstein et al. (2001)). However, lack of information
about exact positions, times, and masses of impacting meteoroids
severely restricts usage of such models. For this reason, as a first
step of interpretation of our observations, we assume local equilibrium of the
exosphere. In this case we can use the sphericalsymmetry model of Chamberlain
(1963) taking into account
changes in gravity with height or, in some approximation, the
simple barometric formula (in our range of altitudes) to obtain
exospheric parameters near the observed point from line-of-sight
column abundances. As a result the scale height H of Na atoms on
August 13/14 is estimated as 700 7200 km using the least-squares
regression of the measured intensities of Na D1 and D2 lines with
theoretical exponential dependence of the intensity on height
(the coefficient of determination $r^{2}$=0.9). The temperature of Na
atoms is estimated as 3100$\pm$800~K from the barometric formula
$T=HA_r(Na)a/R$, where the atomic mass $A_r(Na)=0.023$~kg/mol, $a=1.62$~m/s$^2$
is the gravitational acceleration on the surface of the
Moon, and $R$=8.31~J\,mol$^{-1}$K$^{-1}$ is the gas constant. For observations
of August 12/13 the coefficient of determination $r^2$ is very
low, about 0.1, and the temperature of Na atoms is estimated in the
range between 18,000 and 7000~K. Hence, the temperature of
Na atoms on August 12/13 cannot be estimated because during
observations the column density of Na atoms changes significantly. Meteoroid
bombardment is considered as the main source
of Na atoms at the poles (Sarantos et al., 2010). In this paper the
temperature of Na atoms is estimated to be about 3000~K when
influence of meteoroid bombardment on properties of the lunar
exosphere is maximal, because this time the Moon is located in the
Earth's magnetosphere protecting the lunar surface from the
interaction with solar wind particles. For these reasons, the
temperature of Na atoms on August 12/13, 2009, is assumed to
be the same (3100~K) as that on August 13/14, 2009.

We study several possible models of temporal activity of
the Perseid meteor shower (see Fig. 2). In accordance with
Chamberlain (1963) and assuming the local spherical symmetry
of the exosphere (the case of numerous impacts of small meteoroids) the zenith
column density $N_{zen}$($h$) at the given altitude $h$ is
estimated as

\begin{equation}
N_{zen}=\frac{1}{2}N_{LOS}(h)\frac{K_{zen}(R_{Moon}/H(h))}{K_{tan}(R_{Moon}
/H(h))}
\end{equation}

where $R_{Moon}=1738$~km is the radius of the Moon, $H$($h$) is the scale
height of studied species at the given altitude $h$ in the lunar
atmosphere, $K_{zen}$ and $K_{tan}$ are correction factors to the simple
plane-parallel model of the lunar exosphere with constant gravity
for calculations of $N_{zen}(h)$ and $N_{tan}(h)$, respectively, taken from
Table 3 of Chamberlain (1963). According to Chamberlain (1963),
the surface zenith column density $N_{zen0}$ is calculated as

\begin{equation}
\begin{aligned}
N_{zen0}=N_{zen}(h)\frac{e^{h/H(h)}}{1+(h/R_{Moon})}\frac{\zeta_0(R_{Moon}
/H_0)K_{zen0}(R_{Moon}/H_0)}{\zeta(R_{Moon}/H(h))K_{zen}(R_{Moon}/H(h))}
\end{aligned}
\end{equation}

where $H_0$ is the surface scale height, $\zeta$ is the partition function
taken from Table 2 of Chamberlain (1963), the index ''0'' denotes
the value of these considered functions at the surface. Let us
suppose that the observed increase in the Na surface zenith
column density (see Table 1) on August 12/13, 2009 is caused by
numerous impacts of Perseid meteoroids. In this case the surface
column density of additional impact-produced Na atoms $N_{zenImp0}$
can be estimated as

\begin{equation}
N_{zenImp0} = N_{zen0}(90) - N_{zen0}(270)
\end{equation}
where $N_{zen0}(90)$ is the first spectrum obtained at 90 km $N_{zen0}(270)$ is
the second spectrum obtained at 270 km

Here we use the approximation of the local equilibrium of
the exosphere before and after meteoroid impacts. The $N_{zenImp0}$
value weakly depends on assumed temperature of Na atoms.
Namely, this value is about (4.9$\pm$0.8)${\times}10^8$,
(4.0$\pm$0.6)${\times}10^8$, and
(3.7$\pm$0.5)${\times}10^8$~cm$^{-2}$ at temperature equal to 2000, 3000, and
4000~K, respectively. The found column density of Perseid impact produced Na
atoms at the assumed temperature of 3100~K, about
4${\times}10^8$~cm$^{-2}$, is several times lower than the background column
density of Na atoms above the north pole of the Moon near the last
quarter, about $10^9$~cm$^{-2}$, in agreement with our observations (see
Table 2) and observations of Sprague et al. (2012), and comparable
to the upper limit of the column density of impact-induced Na
atoms delivered to the lunar exosphere by impacts of sporadic
meteoroids, about $3{\times}10^8$~cm$^{-2}$, as estimated by Sarantos et al.
(2008). 

\begin{table*}
\caption{\normalsize Results of spectroscopic search for undetected species in the lunar
exosphere on August 12--14, 2009. The values of g-factors are taken from (a):
Killen et al. (2009), (b): Sarantos et al. (2012), (c): Sullivan and Hunten
(1964), and (d): Flynn and Stern (1996). The dependence of g-factors on radial
velocity is considered for Na 5890 and 5896~\AA, Ca 4227~\AA, Al 3962~\AA, Fe
3859~\AA, and Ti 3990~\AA\ lines; for other lines g-factors are given at zero
Doppler shift. (d) -- column densities are calculated for 400~K at the 5 sigma
level by Flynn and Stern (1996); the depletion factors are estimated for the
case of all mechanisms of delivery of atoms to the exosphere. (e) -- column
densities obtained by Flynn and Stern (1996) were re-calculated for recently
corrected g-values of Si 3906~\AA\ line and Ti 5036~\AA\ line reported by Sarantos
et al. (2012), depletion factors are estimated for the case of all mechanisms of
delivery of atoms to the exosphere. (f) -- Halekas et al. (2013) and (g) --
Poppe et al. (2013), depletion factors are estimated for the case of all
mechanisms of delivery of atoms to the exosphere. (h) -- observations of Cook et
al. (2013) at the assumed temperature of 3000~K and
[Na]=$2.2{\times}10^9$~cm$^{-2}$, depletion factors are estimated for the case
of all mechanisms of delivery of atoms to the exosphere. (i) -- observations
performed on Aug. 12, 23:54 -- Aug. 13, 0:24 UT, 2009. Calculations of
theoretical brightnesses and depletion factors are performed using the following
assumption: $\tau{loss}(K)$=$2{\times}\tau_{bal}(K)$,
$\tau_{loss}(K)$=$2{\times}\tau_{bal}(K)$,
$1/\tau_{loss}(Na)$=$0.9/(2{\times}\tau_{bal}(Na))$+$0.1/\tau_{ion}(Na)$,
$1/\tau_{loss}(Li)$=$0.5/(2{\times}\tau_{bal}(Na))$+$0.5/\tau_{ion}(Li)$, for
atoms of other elements $\tau_{loss}(X)$=$\tau_{bal}(X)$ at 3000~K; for all
elements $F_{unc}(X)=f_{atom}(X)=1$. (k) -- observations performed on August 13,
2009, 23:22--23:52 UT. Depletion factors are estimated for the case of all
mechanisms of delivery of atoms to the exosphere. Calculations of theoretical
brightnesses and depletion factors are performed using the same assumptions as
for the case (i). (l) -- observations performed on Aug. 13, 23:53 -- Aug. 14,
0:23 UT, 2009.\vspace{5mm}}
\renewcommand{\tabcolsep}{3pt}
\renewcommand{\arraystretch}{1.1}
\small

\begin{tabular}{llllllllllll}
\hline
Element & $\lambda$, \AA\ & g-factor, & \multicolumn{2}{c}{Observed brightness Iobs,}& & \multicolumn{3}{c}{Column density,} & Theoretical & \multicolumn{2}{c}{Depletion factor} \\
        &        &(photons           &\multicolumn{2}{c}{(R) (3$\sigma$ level)}&&\multicolumn{3}{c}{cm$^-2$}&brightness&  \multicolumn{2}{c}{relative to Na} \\
\cline{4-5} \cline{7-9} \cline{11-12}
        &        &atom$^{-1}$s$^{-1}$)& & & & & & &$I_{theor}$, $R^h$& \\  
        &        &                    &Aug.12, 23:     &Aug.13, 23:       &&\multicolumn{2}{c}{This work}&$N_{zen0}$,& & Our & Other \\
\cline{7-8}
        &        &                    &13--23:43 UT    &22--23:52 UT & & & & other works & & observations & data \\ 
&&&&&&$N^i_{zenImp0}$&$N^k_{zen0}$&&&&\\
\hline
Na& 5890&0.59$^a$  & 159 & 152 && 4.9$\times$10$^8$   & 9.3$\times$10$^8$   &2.2$\times$10$^9$\,$^d$    &155  & 1    & 1 \\
  &     &       &     &     &&                     &                     &2.3$\times$10$^9$\,$^f$    &     &      &   \\
  &     &       &     &     &&                     &                     &2.2$\times$10$^9$\,$^h$    &     &      &   \\
Na& 5896&0.34$^a$  & 77  & 74  && 4.1$\times$10$^8$   & 7.7$\times$10$^8$   & -                    & 90  & 1    & 1 \\ 
K & 7699&1.94$^b$  & -   & -   && -                   &                     &1.3$\times$10$^9$\,$^d$    & 30  & -    & 0.06$^d$ \\
  &     &       &     &     &&                     &                     &4.2$\times$10$^9$\,$^f$    &     &      & 0.05$^f$ \\ 
Ca& 4227&0.59$^a$  & $<$13 & $<$10 &&$<$1.1$\times$10$^8$   &$<$5$\times$10$^7$     &$<$9.2$\times$10$^7$\,$^d$   & 520 &$>$40i  &$>$150$^d$ \\
  &     &       &     &     &&                     &                     &$<$1.2$\times$10$^9$\,$^f$   &     &$>$100$^k$ &$>$10$^f$ \\
  &     &       &     &     &&                     &                     &$<$6.2$\times$10$^8$\,$^h$   &     &      &$>$22$^h$ \\
Al& 3962&0.035$^b$ & $<$13 & $<$9  &&$<$2$\times$10$^9$     &$<$8.4$\times$10$^8$   &$<$5.1$\times$10$^9$\,$^d$   & 70  & $>$5i  &$>$6$^d$ \\ 
  &     &       &     &     &&                     &                     &$<$2$\times$10$^8$\,$^f$     &     &$>$13$^k$ &$>$150$^f$ \\
  &     &       &     &     &&                     &                     &$<$10$^8$\,$^g$              &     &      &$>$150$^g$ \\
  &     &       &     &     &&                     &                     &$<$6$\times$10$^7$\,$^h$     &     &      &$>$500$^h$ \\
Li& 6708&16$^c$    & $<$30 & $<$24$^k$&&$<$8$\times$10$^6$     &$<$4.9$\times$10$^6$\,$^k$  &$<$1.1$\times$10$^6$\,$^d$   & 11  &$>$0.4i &$>$20$^d$ \\
  &     &       &     &     &&                     &                     &                      &     &$>$1.6$^l$ &     \\
Fe& 3859&0.0041$^b$& $<$15 & $<$15 &&$<$1.9$\times$10$^{10}$&$<$1.1$\times$10$^{10}$&$<$2.5$\times$10$^{10}$\,$^d$&0.9  &$>$0.06i&$>$0.12$^d$\\
  &     &       &     &     &&                     &                     &$<$3.8$\times$10$^{10}$\,$^f$&     &$>$0.1$^k$ &$>$0.08$^f$ \\
  &     &       &     &     &&                     &                     &$<$1.2$\times$10$^{9}$\,$^h$ &     &      &$>$3$^h$ \\
Ba& 5536&11$^d$    & $<$14 & $<$9  &&$<$1.4$\times$10$^7$   &$<$2.2$\times$10$^6$   &$<$7.5$\times$10$^6$\,$^d$   &0.6  &$>$0.04i&$>$0.25$^d$ \\
  &     &       &     &     &&                     &                     &                      &     &$>$0.3$^k$ &      \\
Ti& 3990&0.07$^b$  & $<$20 & $<$15 &&$<$1.4$\times$10$^9$   &$<$6.9$\times$10$^8$   &$<$4.5$\times$10$^{9}$\,$^{d,e}$&0.5 &$>$0.024i&$>$0.02$^{d,e}$\\
  &     &       &     &     &&                     &                     &                      &     &$>$0.05$^k$&  \\
Si& 3906&7.9$\times$10$^{-5}$\,$^b$&$<$17&$<$10&&$<$6.7$\times$10$^{11}$&$<$4.2$\times$10$^{11}$&$<$3.9$\times$10$^{11}$\,$^{d,e}$&0.25&$>$0.014i&$>$0.12$^{d,e}$\\ 
  &     &       &     &     &&                     &                     &$<$5.8$\times$10$^{9}$\,$^f$ &     &$>$0.04$^k$&$>$85$^f$  \\
  &     &       &     &     &&                     &                     &$<$5$\times$10$^{7}$\,$^h$   &     &      &$>$1000$^h$  \\
Mn& 4033&0.0109$^b$& $<$19& $<$11 &&$<$8.9$\times$10$^9$   &$<$3.2$\times$10$^9$   &$<$2.2$\times$10$^9$\,$^h$   &0.04 &$>$0.002i&$>$0.02$^h$\\
  &     &       &     &     &&                     &                     &                      &     &$>$0.006$^k$&  \\
\hline
\end{tabular}
\end{table*}

\begin{figure}
\centering
\includegraphics[width=0.45\textwidth,clip]{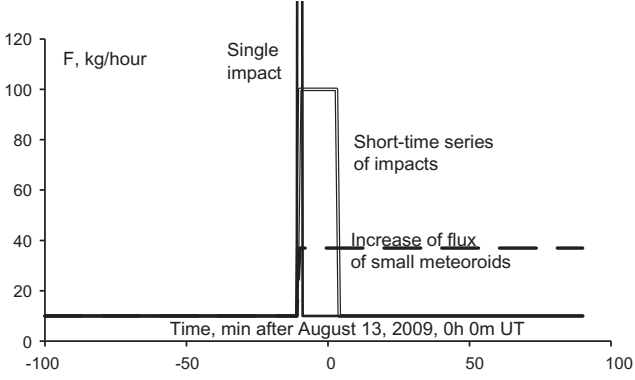} 
\caption{Models of activity of the Perseid meteor shower during observations of
August 12/13, 2009.}
\end{figure}

\begin{table}
\small
\caption{Solar wind density and velocity (ACE-SWEPAM, 2009; OMNI, 2009) and solar F10.7
index measurements (Space weather, 2009) on August 12--14, 2009.}
\renewcommand{\tabcolsep}{3pt}
\centering
\begin{tabular}{llll}
\hline
Date & Solar wind & Solar wind       & Solar F10.7~cm \\
 UT  &   density, & velocity,        & flux, \\
     &(cm$^{-3}$) & (km/s)           & (10$^{-22}$ J s$^{-1}$ m$^{-2}$ Hz$^{-1}$)                 \\
\hline
Aug 12, 23:13 -  & 5$\pm$1 & (340-360)$\pm$3 & 66.5 \\
Aug. 13, 01:13 & & & \\
Aug 13, 23:22 - & ~5$\pm$2& (325-335)$\pm$2 & 66.8 \\
Aug. 14, 00:56  & & & \\
\hline
\end{tabular}
\end{table}

\begin{figure*}
\centering
\includegraphics[width=0.7\textwidth,clip]{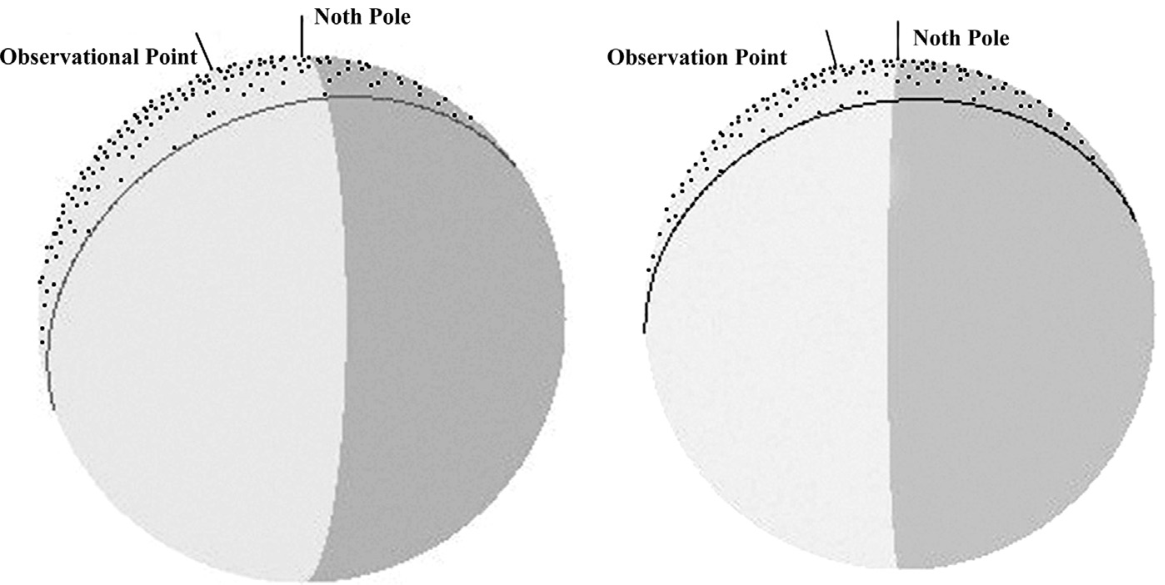} 
\caption{Potential Perseid meteor shower impact region for August 13, 2009, 0 UT (left) and August 14, 2009, 0 UT (right).}
\end{figure*}

Taking into account the global asymmetry of the distribution of
impact-produced Na atoms in the lunar exosphere, the mass of
impact-produced Na atoms $M_{ImpProd}$ is estimated as

\begin{equation}
M_{ImpProd} = \frac{N_{zenImp0obs}A_r(Na) S_{bomb}}{k_{geom} N_A}
\end{equation}

where $S_{bomb}=1.9{\times}10^{17}$~cm$^2$ is the area of the hemisphere
bombarded by the Perseid meteor shower (see Fig. 3), $N_A=6.02{\times}10^{23}$
is the Avogadro constant, $k_{geom}$ is the geometrical factor depending on the
spatial distribution of emitting atoms in the lunar
exosphere. The $k_{geom}$ value is the ratio of the column density
$N_{zenImp0obs}$ at the observed point to the average column density
$N_{zenImp0}$ above the hemisphere bombarded by Perseids:

\begin{equation}
k_{geom}=\frac{2 \pi f_N(b,l)}{\oint f_N(b,l) d\Omega},
\end{equation}

where the model distribution function $f_N(b,l)$ is the surface
column density of impact-produced atoms of considered elements
at given selenographic latitude $b$ and longitude $l$ while the integral
is taken over the whole lunar surface. To model the distribution
function $f_N(b,l)$, we use a simple approximation of homogeneous
meteor shower with large numbers of meteoroid impacts and
small impact-produced clouds. It was assumed that the activity of
the Perseid meteor shower quickly increased soon after recording
of the first spectrum and remained constant during recording of
the second and third spectrum of August 12/13, 2009 (see also
Fig. 2 and Section 4.3 of this paper), because the activity of the
Perseid 2009 meteor shower on Earth might change quickly at the
time scale of about 30 min (IMO, 2009). According to the simulation of
impact-produced lunar exosphere performed by Goldstein
et al. (2001) the surface column density decreases from impact
point sharply, and most of mass of impact-produced matter
concentrates in the small neighborhood of impact points during
first hour after an impact. In this case it is safe to consider the
global distribution function $f_N$ is to be proportional to the frequency of
meteoroid impacts, i.e. to cosine of angular distance $\phi$
from subradiant point on the hemisphere facing the radiant:

\begin{equation}
f_N(\phi)=f_0 cos(\phi)
\end{equation}

and zero on another hemisphere. As a result, $k_{geom}$=2\,cos$\phi_{obs}$,
where $\phi_{obs}$ is the value of angle $\phi$ for the observed point. The
angle $\phi_{obs}$ is equal to 56 and 59$^\circ$; $k_{geom}$ value is about 1.1
and 1.0 for observations of August 12/13 and 13/14, 2009, respectively. By
using Eq. (5), the mass of additional impact-produced Na atoms is
estimated as 3 kg.

We assume this additional mass is produced by increase in the
mass flux of the Perseid meteor shower. To estimate the increase
in the mass flux $\Delta\,F_{imp}$ we can use a simple formula:

\begin{equation}
\Delta\,F_{imp}=\frac{M_{ImpProd}}{\tau_{loss}(Na)}\frac{2}{S_{bomb}}\frac{1}{\{
Na\}(Y+1)},
\end{equation}

where $\tau_{loss}(Na)$ is the lifetime of Na atoms in the exosphere, $S_{bomb}/
2$ is the cross-sectional area of the Moon, $\{Na\}$ is the Na mass fraction in
the impact-induced hot cloud, $Y$ is the target-toimpactor mass ratio in the
cloud. We estimate the lifetime of Na atoms as
$1/\tau_{loss}(Na)=0.9/(2{\times}\tau_{bal}(Na))+0.1/\tau_{ion}(Na)$, where
$\tau_{bal}(Na)=1.3{\times}10^3$~s is the mean time of ballistic flight of Na
atoms at 3100~K, $\tau_{ion}(Na)$ is the Na photoionization lifetime,
$6.2{\times}10^4$~s (Huebner et al., 1992), because we assume that
impact-produced Na atoms can be seen at observed altitudes just during two
ballistic hops (see also Section 4.3 of this paper) while 10\% of Na
atoms at 3100~K are at escaping orbits and lifetimes of escaping
atoms are determined by slow photoionization. According to this
estimation, $\tau_{loss}(Na)=2.88{\times}10^3$~s. The residence time
$\tau_{res}$ of Na atoms at the surface is equal to 0.7 s at 100~K for
adsorption of Na
atoms at quartz and rapidly decreases with increasing temperature (Hunten et
al., 1988). Since $\tau_{res}(Na) << \tau_{loss}(Na)$, we can neglect
the residence time. The north pole of the Moon bombarded by
Perseids has a quite low content of Na-rich KREEP basalts and
gabbronorites and its average elemental composition corresponds
to mixture of ferroan anorthosites and norites with mass ratio of
about 1:1 (Berezhnoy et al., 2005). For this reason the elemental
composition of the lunar regolith is assumed to be the same as the
composition of the mixture of ferroan anorthosites and norites,
namely, 0.28 wt\% for Na (Lodders and Fegley, 1998). The elemental
composition of Perseid meteoroids is assumed to be the same as
composition of CI chondrites, namely, 0.5 wt\% for Na (Lodders and
Fegley, 1998). The target-to-impactor mass ratio $Y$ in the impactinduced hot
cloud is assumed to be 50 for the case of 59 km/s
impacts of Perseid meteoroids in accordance with Cintala (1992).
Thus, the Na mass fraction in the impact-induced hot cloud {Na} is
assumed to be 0.0028. More details about estimation of these
values can be found in Berezhnoy and Klumov (2008) and
Berezhnoy (2013). Finally, the increase in the mass flux $\Delta F_{imp}$
can be estimated with Eq. (8) to be $7.7{\times}10^{-17}$~g\,cm$^{-2}$\,s$^{-1}$
or 27~kg/h. This value corresponds to mass of impacted meteoroids of
about 20~kg during the lifetime of Na atoms in the lunar exosphere, about
2900~s. 

Because $\tau_{loss}(Na)$ is comparable with duration of performed
spectral observations, it is possible to use another model to
describe the sharp increase in Na emission, namely, short-time
series of impacts at about August 13, 2009, 23:50 min UT.

Assuming a homogeneous distribution of impacts along the
section and using Eq. (5), we obtain the same mass
impact-produced Na atoms $M_{ImpProd}=3$~kg, and the
impactors can be estimated by a simple formula:

\begin{equation}
M_{Imp}=\frac{M_{ImpProd}}{\{Na\}(Y+1)}
\end{equation}

Using Eq. (9) we estimate the mass of impactors $M_{imp}$ to be about
20~kg.

Accuracy of our data is not enough to distinguish the model of
flux increase and the model of short-time numerous impacts (see
Fig. 2). However, the nature of both models is the same, therefore
we may consider the estimations of $\Delta F_{imp}$ and $M_{imp}$ as two
different representation of a single phenomenon of Perseid
shower strengthening.

A search for strongest bands of impact-produced diatomic
molecules (namely, the 0--0 band of AlO B$^2\Sigma$--X$^2\Sigma$ (4843 \AA) and
the 0--0 band of MgO B$^1\Sigma^+$--X$^1\Sigma^+$ (4986 \AA)) was also
performed.
Einstein coefficients are equal to $6{\times}10^6$~s$^{-1}$ (Honjou, 2011, 2012)
and $2{\times}10^7$~s$^{-1}$ (Daily et al., 2002) for the considered AlO B--X
and MgO B--X transitions, respectively. Assuming that the duration of
thermal emission is 0.1 s, the electron temperature is 3100~K and
that all Al and Mg are present in the form of AlO and MgO
molecules, intensities of studied transitions are estimated to be
equal to $5{\times}10^{-3}$ and $3{\times}10^{-4}$~R for AlO and MgO,
respectively.
These values are well below the found upper limits, about 1.2~R
for AlO (4843 \AA) and 3~R for MgO (4986 \AA). Thus, even if these
molecules are present in the lunar exosphere, they cannot be
detected with this spectroscopic technique.

A search for other emission lines such as Fe (3860 \AA), Al
(3962 \AA), and Ca (4227 \AA) in recorded spectra was also performed.
Upper limits of the intensitity of expected emission lines are
estimated. Fig. 4 shows the ''non-detection'' of Ca. The same
procedure described in Eqs. (1)--(4) is applied for the analysis of
emission lines of atoms of other elements. It is assumed that the
temperature of atoms of all considered elements is 3100~K. The
found upper limits of abundances of atoms of other elements are
given in Table 2.

\begin{figure}
\centering
\includegraphics[width=0.45\textwidth]{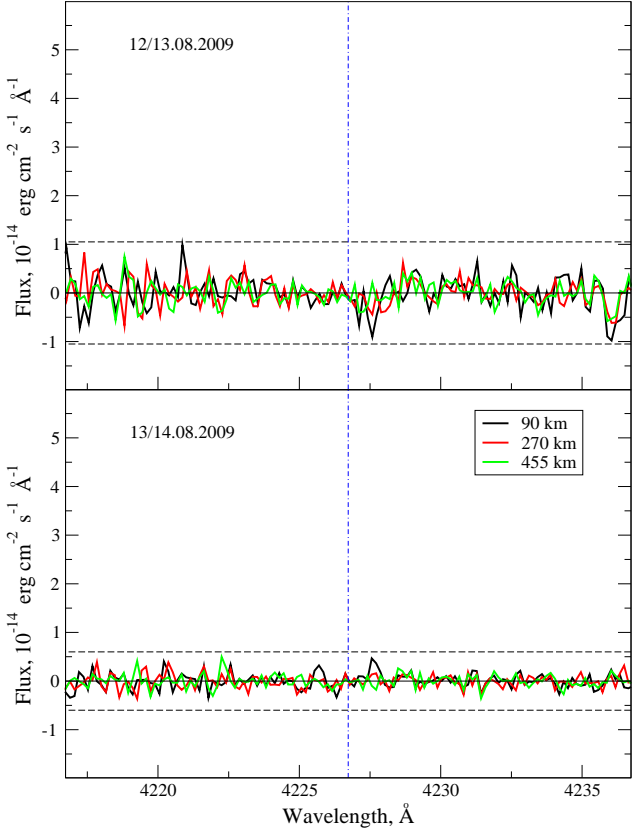} 
\caption{Non-detection of Ca (4227 \AA) line in the lunar exosphere.}
\end{figure}

The theoretical line intensities of atoms of still undetected
elements in the lunar exosphere can be also estimated. In
accordance with Eq. (2) from (Berezhnoy, 2010), the surface zenith
column density $N_{zen0}(X)$ of impact-produced atoms of considered
element X can be calculated as

\begin{equation}
N_{zen0}(X)=\frac{F_{ret}\{X\}\tau_{loss}(X)f_{atom}(X)F_{unc}(X)F_{imp}(Y+1)}{
m(X) }
\end{equation}

where $F_{ret}$ is the retention factor defined as the mass fraction of
the impact-produced cloud captured by the Moon, $\{X\}$ is the mass
fraction of element $X$ in the cloud; $\tau_{loss}(X)$ is the lifetime of atoms
of $X$ in the exosphere; $F_{unc}(X)$ is the fraction of uncondensed
species of element $X$ in the gas phase; $F_{imp}$ is the mass flux of
incoming meteoroids in units of g\,cm$^{-2}$\,s$^{-1}$; and $m(X)$ is the mass
of atoms $X$ in grams. The term $f_{atom}(X)$ is the ratio of the
abundance of atoms of the element $X$ produced directly during
impact and by photolysis to the total abundance of $X$-containing
species in the gas phase of the cloud. In accordance with
Berezhnoy (2013), the retention factor $F_{ret}$ is assumed to be
0.6 and the elemental composition of the impact vapor is assumed
to be that of mixture of CI chondrite, norite, and ferroan anorthosite with mass
ratio equal to 1:25:25. The lifetimes $\tau_{loss}(X)$ of
considered atoms in the exosphere were also estimated in accordance with
Berezhnoy (2013). Using these assumptions and
$F_{unc}(X)=f_{atom}(X)=1$, theoretical brightnesses of atomic lines of
different elements for the case of the Perseid mass flux of 27~kg/h
are estimated and summarized in Table 2.

\section{Discussion}

\subsection{Other sources of Na atoms in the exosphere}

Intensities of solar wind and solar X-rays on August 12/13 and
13/14, 2009 are comparable (Solar monitor, 2009; OMNI, 2009).
Table 3 lists the solar radio flux at 10.7 cm and the solar wind
density and velocity at the time of observations. The last two data
are related to ion sputtering and high values would indicate high
rates; F10.7 index, instead, is a proxy for the solar UV activity,
hence also of the photon stimulated desorption efficiency. All
these values are clearly in the usual ranges. For example, the five minute
average values of the solar wind density during both
nights are roughly constant within 20\% and always below
4~cm$^{-3}$. High-speed streams or highly compressed regions (such
as coronal mass ejections or co-rotating interaction regions) were
not observed at 1 AU, and the space weather was rather calm
during those days (ACE-SWEPAM, 2009). Meanwhile, we cannot
exclude possible influence of some factors in the solar wind on
the discussed phenomenon, which should be investigated in the
future. Changing illumination conditions leads to a higher rate of
photon-simulated desorption of line-of-sight Na atoms from the
surface to the lunar exosphere on August 12/13 as compared to
August 13/14, 2009. However, illumination conditions are changed
too slowly to explain sudden changes of Na column density during
several hours.

Let us assume that activity of the Perseid shower on the Moon
was the same as that on Earth at the same time; in fact, 59~km/s
Perseid meteoroids cross the distance between Earth and the
Moon in about 100 min and the Moon was near the last quarter.
Maxima of the Perseid meteor shower on Earth occurred on
August 12, 2009 at 8 and 18 UT and August 13, 2009, 6 UT.
According to IMO (2009), the activity of Perseid 2009 meteor
shower on Earth in units of zenithal hourly rate (the number of
meteors a single observer would see in one hour under a clear,
dark sky if the radiant of the shower were at the zenith) was equal
to 65 and 25 on August 12, 23 UT -- August 13, 1 UT and August 13,
23 UT -- August 14, 2009, 1 UT, respectively. Thus, we expect a
related enrichment of the content of impact-produced Na atoms in
the lunar exosphere on August 12/13, 2009. Moreover, at the time
of performed observations only the intensity of micrometeorite
bombardment changes significantly, while other sources of the Na
lunar exosphere (ion sputtering and photon-induced desorption)
remain almost constant.

\subsection{Accuracy of estimation of mass of impacted Perseid meteoroids}

Let us estimate the accuracy of our estimation of total mass of
impacting meteoroids. Taking into consideration uncertanties of
the estimation of the temperature of Na atoms, the accuracy of the
Na column density is estimated to be about 50
meteoroids could occur at Na-rich regions. In fact the Na content
in the lunar rocks is variable, from 0.27 wt\% in ferroan anorthosites
and 0.3 wt\% in norites to 0.7 wt\% in gabbronorites and KREEPbasalts (Lodders
and Fegley, 1998). Deposits of Na in the atomic
form are located at the poles of the Moon. About 1.5 kg of Na
atoms were released to the impact-produced plume above the
south pole of the Moon after the LCROSS impact in the crater
Cabeus (Killen et al., 2010). Assuming mass of the plume of about
3000 kg (Colaprete et al., 2010), this amount corresponds to about
0.05 wt\% of Na content at the south pole of the Moon. As a volatile
element, Na is enriched in the impact-produced cloud as compared to that in the
lunar regolith (Gerasimov et al., 1998; Yu and
Hewins, 1998). Both factors may lead to overestimation of mass of
impacted Perseid meteoroids. Additional significant uncertainty
may appear due to poor accuracy of estimations of duration of
Perseid meteoroid peak activity, lifetime of Na atoms in the lunar
exosphere, the target-to-impactor mass ratio in the impactproduced cloud, and
the $k_{geom}$ value. In summary, total mass of
impacted Perseid meteoroids during the lifetime of Na atoms in
the lunar exosphere, about 2900 s, is estimated with an accuracy
of about a factor of three, to the range between 7 and 70 kg. Let us
note that according to estimates of the Perseid mass flux on the
Earth (Hughes and McBride, 1989), the total mass flux of Perseid
meteoroids colliding with the Moon is about 1 kg/h; this value is
significantly lower than our estimate.

\subsection{Single big impact or numerous small impacts?}

To analyze the spectral observations a model of numerous
impacts of small meteoroids (see Eq. (7)) was used. Another
possibility is to model the unique impact of a big meteoroid. The
probability of impact of a 300 kg (80 cm in diameter) sporadic
meteoroid (for this case Y= 3, see Eq. (10)) required to explain the
changes of properties of Na atmosphere of the Moon can be
estimated as the ratio between the duration of our observations
and the frequency of required collisions. Frequency of collision of a
required 80 cm sporadic meteoroid is about one event per week
on Earth (Brown et al., 2002); the probability of such impact
during our observations on August 12/13, 2009 is estimated as low
as $5{\times}10^{-4}$~h$^{-1}$. Based on recent observations of impact events,
this value is estimated as high as $5{\times}10^{-3}$~h$^{-1}$ (Ortiz et al.,
2006; Brown et al., 2013; Madiedo et al., 2014). However, even higher
collision rate of sporadic meteoroids cannot explain observed
significant quick changes of the Na surface zenith column density.
Presence of massive bodies in meteor streams is characterized by
the mass distribution index $s$. Namely, if the $s$ value is less than 2, the
majority of mass is concentrated in several big bodies. In the opposite
case, it is concentrated in small numerious meteoroids. Hypothesis of
the single impact of a big meteoroid is in agreement with average
mass distribution index of 1.45 found through radar observations of
the Perseid showers of 1980--1983, 1985, 1986, 1989, 1991--1993,
1995, 1996, 2000 (Pecinov\'a and Pecina, 2007). Other radar observations give
slighly higher $s$ value equal to 1.61 (\v{S}imek, 1987).
Estimations of the spatial number density of Perseid meteoroids
with diameters of 0.03 mm (Brown and Rendtel, 1996), 1 cm (Beech
and Illingworth, 2001), and 3 m (Smirnov and Barabanov, 1997) and
results of Beech et al. (2004) agree with the Perseid meteor shower
mass distribution index of about 1.6--1.8. Based on Fig. 2 of Beech
et al. (2004) at the assumed $s=1.7$, the frequency of collisions of
30 cm Perseid meteoroids with the Moon is estimated as several
impacts per hour during maximum of the Perseid meteor shower.

However, Apollo seismograms in 1975 detected an impact rate
of small (about 0.1--1 kg) Perseid meteoroids of about 5 events per
day, while large meteoroids (41 kg) were not detected in 1972--
1976 and only the upper limit of the impact rate of such
meteoroids is estimated as about 0.5 events per day (Oberst and
Nakamura, 1991). Only a unique optical flash at the surface of the
Moon caused by Perseid meteoroid is detected (Yanagisawa et al.,
2006). Mass of the impacted meteoroid, about 12~g, and duration
of performed observations agree with low frequency of impacts
of large Perseid meteoroids (41 kg) estimated by Oberst and
Nakamura (1991). The works of Oberst and Nakamura (1991) and
Yanagisawa et al. (2006) can be explained if the $s$ value is about
2.0 or higher. In this case the frequency of 30 cm Perseid
meteoroid impacts can be estimated as low as $3{\times}10^{-3}$~h$^{-1}$.

One of the possible reasons of significant uncertainty of the
impact rate of big Perseid meteoroids is that the $s$ value depends
on time. For example, radar observations give the $s$ value range
between 1.3 and 1.6 during different years Pecinov\'a and Pecina
(2007). These changes may be also responsible for variable impact
rates of massive Perseid meteoroids reported by Oberst and
Nakamura (1991).

Based on our observations, it is possible to try to distinguish
scenarios of the unique impact of big meteoroids, short-time series of
small impacts, and increases in the Perseid mass flux (see Fig. 2).
Namely, after the big unique impact and short-time series of small
impacts, the temperature of impact-produced Na atoms will decrease
quickly while for the case of an increase in the mass flux it remains
constant. Assuming the energy accommodation coefficient $\chi=0.62$
for sodium (Hunten et al., 1988), the surface temperature of 350~K,
and the initial temperature of Na atoms of 3000~K, the effective
temperature will be equal to 1350, 730, and 500~K after a first,
second, and third collision of Na atoms with the surface, respectively,
occurring about 21, 35, and 45 min after the impact, respectively.
These temperatures correspond to scale heights equal to 670, 300,
160, and 110 km, respectively. It may lead to the significant decrease
of the amount of detected Na atoms at 455 km altitude during
recording of the third spectrum if single impact or short-time
bombardment by small meteoroids occurred between August 12,
23:40 UT and August 13, 0:15 UT. Thus, almost equal column
densities of Na atoms estimated from second and third spectra can
be better explained by increase in the mass flux of small meteoroids.
However, if the χ value is significantly lower than 0.6 as discussed by
Smyth and Marconi (1995), then hypotheses of unique impact or
short-time bombardment by small meteoroids are still valid if it
occurred exactly at the middle of the second spectrum of August 12/13, 2009
which seems to be unlikely as discussed above.

\subsection{The presence of atoms of refractory elements in the lunar
exosphere}

In our observations the depletion factor $I_{theor}/I_{obs}$ is defined as
the ratio of predicted theoretical intensity $I_{theor}$ to the observed
upper limits $I_{obs}$ of the emission intensity of the atoms in study at
a given altitude at the 3 sigma level. The estimated lower limits of
depletion factors of refractory elements depends on altitude and
time of performed observations. Namely, for all mechanisms of
delivery of the atoms in study to the lunar exosphere lower limits
of depletion factors were maximal for the first spectrum recorded
on August 13/14, 2009. For the impact delivery, the lower limits of
depletion factors were maximal for the second spectrum recorded
on August 12/13, 2009, because at this period of time the Perseid
meteoroid mass flux was maximal.
The upper limits of Ca and Al column densities are significantly
lower than those estimated using Eq. (10) in assumption that
$F_{unc}(X)=f_{atom}(X)=1$ (see Table 2, observed vs theoretical brightness).
This result can be explained if $F_{unc}(X) f_{atom}(X) << 1$ for Al and Ca.
Depletion of abundances of impact-produced atoms of refractory
elements such as Ca and Al in the lunar exosphere occurs due to
formation of slowly photolyzed molecules and condensation of dust
particles in the cooling impact-produced cloud (Berezhnoy, 2013).
Several other authors also estimated upper limits of refractory
elements in the lunar exosphere (see Table 2). However, these
observations were performed in the absence of major meteor
showers and include information about different sources of atoms
of refractory elements in the lunar exosphere including sporadic
meteoroids and solar photons. Using Eqs. (1) and (4) from
(Berezhnoy, 2013), for the case of all mechanisms of atom delivery
in the lunar exosphere and assuming [Na]=$2.2{\times}10^9$~cm$^{-2}$, lower
limits of depletion factors regarding Na for observations of Flynn and
Stern (1996), Halekas et al. (2013), Cook et al. (2013) are also
estimated (see Table 2). As compared to these results, our observations give
stringent upper limits of the column density for Ca, Ba, and
Ti atoms (see Table 2). Our observations also give a comparable limit
of depletion factors for Ca atoms as compared to the previous best
estimate of Flynn and Stern (1996) while Cook et al. (2013) gave
stronger constraints on behavior of Al, Si and Fe atoms in the lunar
exosphere. As for Li and Ba, our obtained upper limits of the column
densities are comparable with estimated theoretical values while the
theoretical upper limits of the column densities of Fe, Ti, and Mn
atoms are significantly lower than the obtained upper limits even at
the most favorite case of $F_{unc}(X) f_{atom}(X)=1$.

\section{Conclusions}

Increase in the zenith surface column density of Na atoms in the
lunar exosphere on August 13, 2009 at 0--1 UT of about 40
detected while observations of August 13/14, 2009 show the almost
constant surface column density of Na atoms estimated from all
three recorded spectra. Our observations can be better explained by
numerous collisions of small Perseid meteoroids ($<1$~kg) with the
Moon (the mass flux is about 27~kg/h) with respect of single impact
hypothesis (mass of unique meteoroid of about 20~kg). New stringent
upper limits of Ca, Ba, and Ti column densities in the lunar exosphere
are obtained. Lower limits of depletion factors of Ca and Al atoms in
the lunar exosphere during maximum of the Perseid meteor shower
are estimated for the first time and explained by formation of silicate
dust particles during cooling of an expanded impact-produced cloud.
Our observations confirm influence of the meteoroid bombardment
on the properties of the lunar exosphere. More observations of the
lunar exosphere during activity of meteor showers are required for
deeper understanding of these phenomena.

\section*{Acknowledgments}
Authors thank S.F. Velichko (Terskol Observatory) for performing the
observations and T.K. Churyumova, A.Ya. Freidzon, O.V. Khabarova, A.M. Mozgova,
P. Pecina, and anonymous referee for helpful suggestions.





\section*{References}

ACE-SWEPAM, 2009, (http://www.srl.caltech.edu/ACE/ ASC/level2/index.html) 
Barbieri, C., Benn, C.R., Cremonese, G., Verani, S., Zin, A., 2001. Meteor
showers on the lunar atmosphere. Earth Moon Planets 85--86, 479--486.

Beech, M., Illingworth, A., 2001. 2001 Perseid fireball observations. WGN, J.
Int. Meteor Org. 29, 181--184.

Beech, M., Illingworth, A., Brown, P., 2004. A telescopic search for large
Perseid meteoroids. Mon. Not. R. Astron. Soc. 348, 1395--1400.

Berezhnoy, A.A., Hasebe, N., Kobayashi, M., Michael, G.G., Okudaira, O.,
Yamashita, N., 2005. A three end-member model for petrologic analysis of lunar
prospector gamma-ray spectrometer data. Planet. Space Sci. 53, 1097--1108.

Berezhnoy, A.A., Klumov, B.A., 2008. Impacts as a source of the atmosphere on
Mercury. Icarus 195, 511--522.

Berezhnoy, A.A., 2010. Meteoroid bombardment as a source of the lunar exosphere.
Adv. Space Res. 45, 70--76.

Berezhnoy, A.A., 2013. Chemistry of impact events on the Moon. Icarus 226,
205--211.

Brown, P., Rendtel, J., 1996. The Perseid meteoroid stream: characterization of
recent activity from visual observations. Icarus 124, 414--428.

Brown, P., Spalding, R.E., ReVelle, D.O., Tagliaferri, E., Worden, S.P., 2002.
The flux of small near-Earth objects colliding with the Earth. Nature 420,
294--296.

Brown, P.G., Assink, J.D., Astiz, L., Blaauw, R., Boslough, M.B., Borovi\v{c}ka,
J., Brachet, N., Brown, D., Campbell-Brown, M., Ceranna, L., Cooke, W., de
Groot-Hedlin, C., Drob, D.P., Edwards, W., Evers, L.G., Garces, M., Gill, J.,
Hedlin, M., Kingery, A., Laske, G., Le Pichon, A., Mialle, P., Moser, D.E.,
Saffer, A., Silber, E., Smets, P., Spalding, R.E., Spurný, P., Tagliaferri, E.,
Uren, D., Weryk, R.J., Whitaker, R., Krzeminski, Z, 2013. A 500-kiloton airburst
over Chelyabinsk and an enhanced hazard from small impactors. Nature 503,
238--241.

Burger, M.H., Killen, R.M., McClintock, W.E., Vervack Jr., R.J., Merkel, A.W.,
Sprague, A.L., Sarantos, M., 2012. Modeling MESSENGER observations of calcium in
Mercury's exosphere. J. Geophys. Res. 117 (CiteID E00L11) (15 pp.),
http://dx.doi.org/10.1029/2012JE004158

Chamberlain, J.W., 1963. Planetary coronae and atmospheric evaporation. Planet
Space Sci. 11, 901--960.

Chamberlain, J.W., Hunten, D.M., 1987. Theory of Planetary Atmospheres. Academic
Press, Orlando p. 481

Churyumov, K.I., Berezhny, O.O., Ponomarenko, V.O., Baransky, O.R., Churyumova,
T.K., Kleshchenok, V.V., Mozgova, A.M., Kovalenko, N.S., Shevchenko, V.V.,
Kozlova, E.A., Pakhomov, Yu.V., Velikodsky, Yu.I., 2012. Observations of the
non-stationary atmosphere of the Moon and some its parameters. Astron.
School's Rep. 8, 175--181 (in Ukrainian).

Cintala, M., 1992. Impact-induced thermal effects in the lunar and Mercurian
regoliths. JGR 97, 947--973.

Colaprete, A., Schultz, P., Heldmann, J., Wooden, D., Shirley, M., Ennico, K.,
Hermalyn, B., Marshall, W., Ricco, A., Elphic, R.C., Goldstein, D., Summy, D.,

Bart, G.D., Asphaug, E., Korycansky, D., Landis, D., Sollitt, L., 2010.
Detection of water in the LCROSS ejecta plume. Science 330, 463--468.
Cook, J.C., Stern, S.A., Feldman, P.D., Gladstone, G.R., Retherford, K., Tsang,
C.C.C., 2013. New upper limits on numerous atmospheric species in the native
lunar atmosphere. Icarus 225, 681--687.

Daily, J.W., Dreyer, C., Abbud-Madrid, A., Branch, M.C., 2002. Transition
probabilities in the B$^1\Sigma^+$--X$^1\Sigma^+$ and the
B$^1\Sigma^+$--A$^1\Pi$ electronic systems of MgO. J. Mol. Spectrosc. 214,
111--116.

Flynn, B., Mendillo, M., 1993. A picture of the moon's atmosphere. Science 261,
184--186.

Flynn, B.C., Stern, S.A., 1996. A spectroscopic survey of metallic species
abundances in the lunar atmosphere. Icarus 124, 530--536.

Gerasimov, M.V., Ivanov, B.A., Yakovlev, O.I., Dikov, Yu.P., 1998. Physics and
chemistry of impacts. Earth Moon Planets 80, 209--259.

Goldstein, D.B., Austin, J.V., Barker, E.S., Nerem, R.S., 2001. Short-time
exosphere evolution following an impulsive vapor release on the Moon. J.
Geophys. Res. 106 (E12), 32841--32845.

Halekas, J.S., Poppe, A.R., Delory, G.T., Sarantos, M., McFadden, J.P., 2013.
Utilizing ARTEMIS pickup ion observations to place constraints on the lunar
atmosphere. J. Geophys. Res. 118, 81--88.

Honjou, N., 2011. Ab initio study of band strengths for the
F$^2\Sigma^+$--A$^2\Pi$ electronic transition of AlO. Comput. Theoret. Chem.
978, 138--142.

Honjou, N., 2012. (private communication).

Huebner, W.F., Keady, J.J., Lyon, S.P., 1992. Solar photo rates for planetary
atmospheres and atmospheric pollutants. Astrophys. Space Sci. 195 (1--289),
291--294.

Hughes, D., McBride, N., 1989. The mass of meteoroid streams. Mon. Not. R.
Astron. Soc. 240, 73--79.

Hunten, D.M., Shemansky, D.E., Morgan, T.M., 1988. The Mercury atmosphere. In:
Vilas, F., Chapman, C., Matthews, M. (Eds.), Mercury. University of Arizona
Press, Tucson, AZ, pp. 562--612

Hunten, D.M., Cremonese, G., Sprague, A.L., Hill, R.E., Verani, S., Kozlowski,
R.W.H., 1998. The Leonid meteor shower and the lunar sodium atmosphere. Icarus
136, 298--303.

Hunten, D.M., Kozlowski, R.W.H., Sprague, A.L, 1991. A possible meteor shower on
the Moon. Geophys. Res. Lett. 18, 2101--2104.

IMO, 2009. Perseids 2009: Visual Data Quicklook.
(http://www.imo.net/live/perseids2009)

Killen, R.M., Shemansky, D., Mouawad, N., 2009. Expected emission from Mercury's
exospheric species, and their ultraviolet--visible signatures. Astrophys. J.
Suppl.Ser. 181, 351--359.

Killen, R.M., Potter, A.E., Hurley, D.M., Plymate, C., Naidu, S., 2010.
Observations of the lunar impact plume from the LCROSS event. Geophys. Res.
Lett. 37 (CiteID L23201).

Lodders, K., Fegley, B., 1998. The Planetary Scientist Companion. Oxford
University Press, New York, p. 371

Madiedo, J.M., Ortiz, J.L., Morales, N., Cabrera-Caño, J., 2014. A large lunar
impact blast on 2013 September 11. Mon. Not. R. Astron. Soc. 439, 2364--2369.

Mendillo, M., Baumgardner, J., Flynn, B., 1991. Imaging observations of the
extended sodium atmosphere of the Moon. Geophys. Res. Lett. 18, 2097--2100.

Mendillo, M., Flynn, B., Baumgardner, J., 1993. Imaging experiments to detect an
extended sodium atmosphere on the Moon. Adv. Space Res. 13, 313--319.

Mendillo, M., Baumgardner, J., Wilson, J., 1999. Observational test for the
solar wind sputtering origin of the Moon's extended sodium atmosphere. Icarus
137, 13--23.

Oberst, J., Nakamura, Y., 1991. A search for clustering among the meteoroid
impacts detected by the Apollo lunar seismic network. Icarus 91, 315--325.

OMNI, 2009. (http://omniweb.gsfc.nasa.gov) Ortiz, J.L., Aceituno, F.J., Quesada,
J.A., Aceituno, J., Fern\'andez, M., Santos-Sanz, P., Trigo-Rodr\'iguez, J.M.,
Llorca, J., Mart\'in-Torres, F.J., Monta\~{n}\'es-Rodr\'iguez, P., Pall\'e, E.,
2006. Detection of sporadic impact flashes on the Moon: implications for the
luminous efficiency of hypervelocity impacts and derived terrestrial impact
rates. Icarus 184, 319--326.

Pecinov\'a, D., Pecina, P., 2007. Radar meteors range distribution model. II.
Shower flux density and mass distribution index. Contrib. Astron. Obs. Skaln.
Pleso 37, 107--124.

Poppe, A.R., Halekas, J.S., Samad, R., Sarantos, M., Delory, G.T., 2013.
Model-based constraints on the lunar exosphere derived from ARTEMIS pickup ion
observations in the terrestrial magnetotail. J. Geophys. Res. 118, 1135--1147.

Potter, A.E., Morgan, T.H., 1988. Discovery of sodium and potassium vapor in the
atmosphere of the Moon. Science 241, 675--680.

Sarantos, M., Killen, R.M., Sharma, A.S., Slavin, J.A., 2008. Influence of
plasma ions on source rates for the lunar exosphere during passage through the
Earth's magnetosphere. Geophys. Res. Lett. 35 (CiteID L04105).

Sarantos, M., Killen, R.M., Sharma, A.S., Slavin, J.A., 2010. Sources of sodium
in the lunar exosphere: modeling using ground-based observations of sodium
emission and spacecraft data of the plasma. Icarus 205, 364--374.

Sarantos, M., Killen, R.M., McClintock, W.E., Bradley, E.T., Vervack, R.J.,
Benna, M., Slavin, J.A., 2011. Limits to Mercury's magnesium exosphere from
MESSENGER second flyby observations. Planet. Space Sci. 59, 1992--2003.

Sarantos, M., Killen, R.M., Glenar, D.A., Benna, M., Stubbs, T.J., 2012.
Metallic species, oxygen and silicon in the lunar exosphere: upper limits and
prospects for LADEE measurements. J. Geophys. Res. 117 (CiteID A03103).

\v{S}imek, M., 1987. Dynamics and evolution of the structure of five meteor
streams. Bull. Astron. Inst. Czech. 38, 80--91.

Smirnov, M.A., Barabanov, S.I., 1997. The optical observations of meteoroids in
nearEarth space. In: Proceedings of 2nd European Conference on Space Debris. ESA
SP-393, ESOC, Darmstadt, Germany. pp. 155--157.

Smith, S.M., Wilson, J.K., Baumgardner, J., Mendillo, M., 1999. Discovery of the
distant lunar sodium tail and its enhancement following the Leonid meteor
shower of 1998. Geophys. Res. Lett. 26, 1649--1652.

Smyth, W.H., Marconi, M.L., 1995. Theoretical overview and modeling of the
sodium and potassium atmospheres of the Moon. Astrophys. J. 443, 371--392.

Solar Monitor, 2009. (http://www.solarmonitor.org) 

Space Weather, 2009.
(http://www.spaceweather.ca/data-donnee/sol\_flux/sx-5eng.php)

Sprague, A.L., Hunten, D.M., Kozlowski, R.W.H., Grosse, F.A., Hill, R.E.,
Morris, R.L., 1998. Observations of sodium in the lunar atmosphere during
International Lunar Atmosphere Week, 1995. Icarus 131, 372--381.

Sprague, A.L., Sarantos, M., Hunten, D.M., Hill, R.E., Kozlowski, R.W.H., 2012.
The lunar sodium atmosphere: April--May 1998. Can. J. Phys. 90, 725--732.

Stern, S.A., Parker, J.W., Morgan, Th.H., Flynn, B.C., Hunten, D.M., Sprague,
A., Mendillo, M., Festou, M.C., 1997. NOTE: an HST search for magnesium in the
lunar atmosphere. Icarus 127, 523--526.

Sullivan, H.M., Hunten, D.M., 1964. Lithium, sodium and potassium in the
twilight airglow. Canadian Journal of Physics 42, 937--956.

Tug, H., 1977. Vertical extinction on La Silla. ESO Messenger 11, 7--8.
Verani, S., Barbieri, C., Benn, C., Cremonese, G., 1998. Possible detection of
meteor stream effects on the lunar sodium atmosphere. Planet. Space Sci. 46,
1003--1006.

Verani, S., Barbieri, C., Benn, C.R., Cremonese, G., Mendillo, M., 2001. The
1999 Quadrantids and the lunar Na atmosphere. Mon. Not. R. Astron. Soc. 327,
244--248.

Yanagisawa, M., Ohnishi, K., Takamura, Y., Masuda, H., Sakai, Y., Ida, M.,
Adachi, M., Ishida, M., 2006. The first confirmed Perseid lunar impact flash.
Icarus 182, 489--495.

Yu, Y., Hewins, R.H., 1998. Transient heating and chondrule formation: evidence
from sodium loss in flash heating simulation experiments. Geochim. Cosmochim.
Acta 62, 159--172.

\end{document}